\documentstyle[12pt]{article}
\newcommand{\al}{\alpha}
\newcommand{\be}{\beta}\newcommand{\vr}{\varrho}
\newcommand{\de}{\delta}

\newcommand{\ve}{\varepsilon}\newcommand{\la}{\lambda}

\newcommand{\fr}[2]{\frac{\partial{#1}}{\partial{#2}}}

\newcommand{\lf}{\left}\newcommand{\ri}{\right}
\newcommand{\si}{\sigma}
\newcommand{\vp}{\varphi} \newcommand{\qu}{\quad}
\newcommand{\bib}{\bibitem}\newcommand{\ci}{\cite}
\newcommand{\ra}{\rightarrow}
\newcommand{\ti}{\tilde}\newcommand{\De}{\Delta}
\newcommand{\ta}{\tau}
\newcommand{\equ}{\equiv}\newcommand{\ha}{\hat}
\newcommand{\mb}{\mbox}\newcommand{\th}{\theta}
\newcommand{\bs}{\begin{sloppypar}} \newcommand{\es}{\end{sloppypar}}
\newcommand{\beq}{\begin{equation}} \newcommand{\eeq}{\end{equation}}
\newcommand{\ka}{\kappa}\newcommand{\p}[1]{(\ref{#1})}
\newcommand{\lab}{\label} \newcommand{\Om}{\Omega}

\begin {document}
\begin{center}
\bf MULTI-GAP SUPERFLUIDITY IN NUCLEAR MATTER
\\
\rm A.I. Akhiezer,  A.A. Isayev,
 S.V. Peletminsky,  A.A.  Yatsenko
\\  {\it Kharkov Institute of Physics and Technology,\\
 Academicheskaya Str.
1, Kharkov, 310108, Ukraine} \end{center}

\begin{abstract}
     It is shown that under lowering  density or temperature a
nucleon Fermi superfluid can undergo a phase transition to a new superfluid
state corresponding to superposi\-tion of states with singlet-triplet (ST) and
triplet-singlet (TS) pairing of nucleons (in spin and isospin spaces). Such
states  arise as a result of branching from one-gap solution of the
self-consistent equations, describing ST pairing of nucleons. The density and
temperature  dependence of the order parameters for  new two-gap solutions
is determined in the model with Skyrme effective forces.
\end{abstract}


\vspace{3mm}

As it is known, at sufficiently low temperatures a Fermi-liquid (FL) becomes
unstable with respect to  formation of Cooper pairs and passes into a
superconducting (superfluid) state. Corresponding examples are an
electron liquid in metals  and  superfluid phases of $^3$He.
There are some experimental facts in favour of superfluidity of atomic
nucleus as well~\ci{S}. Since the study of superfluidity of atomic nuclei is
 difficult due to  finiteness of their sizes, it is of interest to study
 superfluidity of infinite nuclear matter, especially in view of
astrophysical applications.  This question has been considered in many works.
However, earlier the normal-to-superfluid phase transition in nuclear matter
has been studied only. In principle, at further lowering of  temperature a
superfluid FL of nucleons can again become unstable  passing into a new
superfluid state.  Such states are characterized by not by one but a few
order parameters and have a lower symmetry than the initial superfluid state.
Thus, we are speaking of the superfluid-to-superfluid phase transitions in
nuclear matter.  We shall clarify the mechanism of appearance of new
superfluid states as well as find the temperature dependence of the order
 parameters for these states.

{\bf 1. Basic equations.} We study
multi-gap superfluidity of nuclear matter using Landau's semiphenomenological
concept of a Fermi-liquid. The basic formalism is laid out in more detail
in Ref. \ci{AIP}, where the phase transitions to one-gap superfluid states of
symmetrical nuclear matter have been studied. The present work is an
extension of Ref.~\ci{AIP} and its main aspect is analysis of the phase
transitions to multi-gap superfluid states. For simplicity we assume that the
energy functional is invariant under rotations in the configurational, spin
and isospin spaces (for infinite uniform nuclear matter the spin-orbit
interaction is equal to zero).  Hence, superfluid phases (Cooper pairs) are
classified by specifying the total spin of the pair, $S=0,1$, the isospin
$T=0,1$, their projections $S_z$ and $T_z$ on the $z$ axis, and the orbital
angular momentum $L=0,1,2,...$ The possible values of the orbital angular
momentum $L$ for each value of $S$ and $T$ must be found according to the
Pauli principle.  Each superfluid phase is described by its own set of order
parameters: at $S=0,\, T=0$ by the scalar order parameter $\De_{00}$
(singlet-singlet pairing of nucleons), at $S=1,\, T=0$ or $S=0,\, T=1$ by the
vector order parameter $\De_{k0}$ or $\De_{0k}$, respectively ($k=1,2,3$) (TS
and ST pairing of nucleons) and at $S=1,\, T=1$ by the tensor order parameter
$\De_{ik}$ ($i,k=1,2,3$) (triplet-triplet pairing of nucleons).  A nucleon
superfluid FL is described by two distribution functions: the normal
distribution function $f_{\ka_1\ka_2}=\mb{Tr}\,\vr a^+_{\ka_2}a_{\ka_1}$ and
the anomalous distribution function $g_{\ka_1\ka_2}=\mb{Tr}\,\vr
a_{\ka_2}a_{\ka_1}$, where $a^+_\ka$ and $a_\ka$ are the creation and
annihilation operators of fermions with momentum $\vec p$, spin (isotopic
spin) projection $\si(\ta)$, $\ka\equ(\vec p,\si,\tau)$, $\mb{Tr}\vr...$
being the average of the operators with the density matrix of the system,
$\vr$. The energy of the system is specified as a functional of the
distribution functions $f$ and $g$, $E=E(f,g)$. It determines the fermion
one-particle energy $\ve$ and the matrix order parameter $\De$ of the system
\beq \ve_{\ka_1\ka_2}=\fr{E}{f_{\ka_2\ka_1}},\qu
\De_{\ka_1\ka_2}=2\fr{E}{g_{\ka_2\ka_1}^+} \label{1}\eeq
The matrix self-consistent equation for determining the distribution
functions $f$ and $g$ follows from the minimum condition for the
thermodynamic potential $\Om=-S+Y_0E+Y_{4a}N_a+Y_{4b}N_b$ ($S$ is the
entropy, $N_a, N_b$ are the numbers of particles of species $a$ and $b$,
$Y_0=1/T,\ Y_{4a}=-\mu_a/T$ and $Y_{4b}=-\mu_b/T$ are the Lagrange
multipliers, $\mu_a$ and $\mu_b$ are the chemical potentials of particles of
species $a $ and $b$, $T$ is the temperature):
\beq
\hat f=\lf\{\mbox{exp}(Y_0\hat\ve+\ha Y_4)+1\ri\}^{-1}\equ
\lf\{\mbox{exp}(Y_0\hat\xi)+1\ri\}^{-1},\lab{2}\eeq
$$
\hat
f=\lf(\begin{array}{cc}f&g\\g^+&1-\ti f\end{array}\ri),\;
\ha\ve=\lf(\begin{array}{cc}\ve&\De\\ \De^+&
-\ti\ve\end{array}\ri),\;
\ha Y_4=\lf(\begin{array}{cc}Y_4&0\\ 0&
-Y_4\end{array}\ri)
$$
Here  the quantities $\ve,\De,Y_4$ are, in turn, matrices in the space of the
$\ka$ variables, with $Y_{4\ka_1\ka_2}=Y_{4\ta_1}\de_{\ka_1\ka_2}$
$(\ta_1=a,b)$ and the tilde stands for the transposition operation. Using
the procedure of block diagonalization \ci{AIP,AKP} in an extended space
$2\times2$, where the operator $\ha f$ acts, one can  express  evidently the
distribution functions $f$ and $g$ in terms of the quantities $\ve$ and
$\De$.  In what
follows, we consider the case of symmetrical nuclear matter (equal  numbers
of protons and neutrons, or $Y_{4a}=Y_{4b}\equ Y_4,\,\, \mu_a=\mu_b\equ\mu$).
In this case we obtain \beq
f=\frac{1}{2}\lf(1-\frac{\xi}{E}\tanh\frac{Y_0E}{2}\ri),\;
g=-\frac{1}{2E}\tanh\frac{Y_0E}{2}\cdot\De;\qu
E=\sqrt{\xi^2+\De\De^+},\;
\xi=\ve+\frac{Y_4}{Y_0} \lab{3}\eeq
We emphasize that the quantities $f,g,\xi$ and $\De$ in Eq.\p{3} are matrices
in the spin, isospin and momentum spaces.
Further we shall
study the unitary states of superfluid nuclear matter, for which the
product $\De\De^+$ is proportional to the product of the unit matrices in the
spin and isospin spaces, $\De\De^+\propto \si_0\tau_0$.  We
consider as a particular case  two-gap unitary states, for which the
order parameters possess by the same symmetry properties on momentum:
\beq
\De({\vec p})=\De_{30}(\vec p)\si_3\si_2\ta_2+
\De_{03}(\vec p)\si_2\ta_3\ta_2,\qu
\De_{30}(-\vec p)=\De_{30}(\vec p),\;  \De_{03}(-\vec p)=\De_{03}(\vec p)
 \lab{7}\eeq
In this case the wave function of a Cooper pair describes the superposition of
states with the triplet-singlet and singlet-triplet pairing of nucleons (TS-ST
states):
\beq
g({\vec p})=g_{30}(\vec p)\si_3\si_2\ta_2+
g_{03}(\vec p)\si_2\ta_3\ta_2,\qu
g_{30}(-\vec p)=g_{30}(\vec p),\;  g_{03}(-\vec p)=g_{03}(\vec p),
\lab{8}\eeq
with the projections of total spin and isospin $S_z=T_z=0$.

{\bf 2. Self-consistent equations. The case T=0}.
To derive the self-consistent equations of a nucleon superfluid FL it is
necessary to specify the energy functional, which we set in the form
 \begin{equation}
 E(f,g)=E_0(f)+E_{int}(g),\;\qu
{E}_0(f)=\frac{4}{V}\sum\limits_{ \vec p}^{}
\ve_0(\vec p)f_{00}(\vec p),\; \ve_0(\vec p)=\frac{\vec p^{\,2}}{2m},
\lab{9}
 \end{equation}
$${E}_{int}(g)=\frac{2}{V}\sum\limits_{ \vec p,\vec q}^{}
\{g_{30}^*( \vec p)V_1( \vec p, \vec q)g_{30}(\vec q) +
g_{03}^*(\vec p)V_2(\vec p,\vec q)g_{03}(\vec q)\}$$
Here $m$ is the effective nucleon mass, $f_{00}$ is the coefficient of the
product $\ta_0\si_0$ in expansion of the distribution function $f$ in the
Pauli matrices, $V_1,V_2$ are the anomalous FL interaction amplitudes, which
have the symmetry properties
$V_i(-\vec p,\vec q)=V_i(\vec p,-\vec q)=V_i(\vec p,\vec q),\, i=1,2.
$
The quantity $m$ contains  account of the normal FL effects and represents
itself the mass of a free nucleon, renormalized by the normal FL
interaction (in the Skyrme model the quantity $m$ is given by Eq.
\p{14}). With allowance for Eqs. \p{1}, \p{9}, we obtain expressions for the
quantities $\De_{30},\De_{03}$ as functions of the quantities $g_{30}$ and
$g_{03}$:
\beq \De
_{30}(\vec p)=\frac {1}{ V}\sum\limits_{\vec q}^{} V_1(\vec p,\vec q
)g_{30}(\vec q), \qu  \De _{03}(\vec p)=\frac {1}{
V}\sum\limits_{\vec q}^{} V_2(\vec p,\vec q)g_{03}(\vec q) \lab{10}\eeq
Conversely, expressions for the distribution functions $g_{03}$ and $g_{30}$
as functions of the quantities $\xi, \De_{30},\De_{03}$ can be found from
 Eqs.  \p{3}, \p{9}. With account of them,
we get
the self-consistent equations for determining the order
parameters of two-gap unitary states of superfluid nuclear
matter:
\beq
\De
_{30}(\vec p)=-\frac {1}{ V}\sum\limits_{\vec q}^{} V_1(\vec p,\vec q
)\lf\{
\frac{\De_+(\vec q)}{4E_+(\vec q)}\tanh\frac{Y_0E_+(\vec q)}{2}+
\frac{\De_-(\vec q)}{4E_-(\vec q)}\tanh\frac{Y_0E_-(\vec q)}{2}\ri\},
\lab{12}\eeq
$$\De
_{03}(\vec p)=-\frac {1}{ V}\sum\limits_{\vec q}^{} V_2(\vec p,\vec q
)\lf\{
\frac{\De_+(\vec q)}{4E_+(\vec q)}\tanh\frac{Y_0E_+(\vec q)}{2}-
\frac{\De_-(\vec q)}{4E_-(\vec q)}\tanh\frac{Y_0E_-(\vec q)}{2}\ri\}$$
Here $\De_\pm=\De_{30}\pm\De_{03},\,E_\pm=\sqrt{\xi^2+|\De_\pm|^2}$.
According to Eq. \p{12}, considering a nucleon superfluid FL   is
characterized by two types of  fermion excitations with the gaps $\De_\pm$
in the spectrum.
Eqs. \p{12} generalize  equations of the BCS theory and contain the
one-gap solutions with $\De_{30}\not=0,\,\De_{03}\equ0$  (TS
pairing) and $\De_{30}\equ0,\,\De_{03}\not=0$ (ST pairing) as
particular cases.

Let us give  an analysis of Eqs. \p{12}, using the model representations of
the BCS theory. Precisely, we assume that the interaction amplitudes $V_1$
and
$V_2$ are not equal to zero only in a narrow layer near the Fermi surface:
$|\xi|\leq\theta,\;\theta\ll\ve_F$ (in numerical calculations we set
$\theta=0.1\ve_F$).
                           Besides, we choose the effective Skyrme forces
\ci{BGH} as the amplitudes of NN interaction. Relation between the anomalous
FL amplitudes  $V_1$, $V_2$ and the amplitudes of NN interaction has been
set in Ref. \ci{AIP}. Taking into account it, for the quantities
$V_1,V_2$ we have:  \beq V_{1,2}(\vec p,\vec
q)=t_0(1\pm x_0)+\frac{1}{6}t_3n^\al(1\pm x_3)+\frac{1}{2\hbar^2}t_1(1\pm
x_1)(\vec p^{\,2}+\vec q^{\,2}),\lab{13}\eeq where $n$ is the density of
symmetrical nuclear matter, $t_i,x_i,\al$ are some phenomenological
parameters. Note that the amplitudes $V_1,V_2$ contain no dependence
from the parameters $t_2,x_2$, because the amplitudes $V_1,V_2$ are the even
functions of the arguments $\vec p,\vec q$ (see details in Ref.
\ci{AIP}). There are sets of parameters $t_i,x_i,\al$, which differ for
various versions of the Skyrme forces (we shall use the Ska, SkM, SkM$^*$
and RATP potentials \ci{BGH} as well as the  SkP  potential \ci{DFT}). In
the Skyrme model the effective nucleon mass is given by the expression \beq
\frac{\hbar^2}{2m}=\frac{\hbar^2}{2m_0}+\frac{1}{16}(3t_1+t_2(5+
4x_2))n,\lab{14}\eeq
$m_0$ being the mass of a  free nucleon.

Let us consider the case $T=0$. As  a result of the above assumptions, we
arrive at  equations for determining the quantities
$\De_{30}\equ\De_{30}(p=p_F),\,\De_{03}\equ\De_{03}(p=p_F)$:
\beq \De_{30}=g_1\int_{-\th}^\th d\xi
\{(\frac{1}{E_+}+\frac{1}{E_-})\De_{30}+
(\frac{1}{E_+}-\frac{1}{E_-})\De_{03}
\},\lab{15}\eeq
$$\De_{03}=g_2\int_{-\th}^\th d\xi
\{(\frac{1}{E_+}+\frac{1}{E_-})\De_{03}+
(\frac{1}{E_+}-\frac{1}{E_-})\De_{30}
\}$$
Here
$\displaystyle{ g_{1,2}=-
\frac{\nu_FV_{1,2}(p=p_F,q=p_F)}{4}}$ and
$\displaystyle{\nu_F=\frac {mp_F}{2\pi^2\hbar^3} }$ being the density of
states at the Fermi surface of a proton (neutron) with the given spin
projection. Eqs.  \p{15} allow to determine the order parameters
$\De_{30},\De_{03}$ in the TS--ST states of paired nucleons. In general case
this can be done only numerically. Analytical consideration  is possible if
the conditions $\De_{30},\De_{03}\ll\th $ are fulfilled (logarithmic
approximation).  Introducing the ratio of the order parameters
$\al=\De_{30}/\De_{03}$ and performing integration in Eqs. \p{15} in this
approximation, we arrive at the following equations for the quantities $\al$
and $\De_{03}$: \beq
\frac{1}{g_1}=4\ln\frac{2\th}{\De_{03}}+2(\frac{1}{\al}\ln
\lf|\frac{1-\al}{1+\al}
\ri| -\ln|1-\al^2|),\lab{16}\eeq
$$\frac{1}{g_2}=4\ln\frac{2\th}{\De_{03}}+
2(\al\ln
\lf|\frac{1-\al}{1+\al}
\ri| -\ln|1-\al^2|).$$
Excluding  $\De_{03}$ from Eqs. \p{16}, we obtain the equation
\beq\vp(\al)=\frac{1}{g_1}-\frac{1}{g_2},\qu \vp(\al)\equ
2(\al-\frac{1}{\al})\ln\lf|\frac{1+\al}{1-\al}\ri|\lab{17}\eeq
Since  $\vp_{min}=\vp(0)=-4$ (the point $\al=0$ is the point of  removable
discontinuity) and $\vp_{max}=\vp(\pm\infty)=4$, then Eq. \p{17} has the
solution for $\al$, if the coupling constants $g_1$ and $g_2$ satisfy
the unequalities
 $
-4<1/g_1-1/g_2<4.$
These unequalities have to be fulfilled  in the
logarithmic approximation for existence of the TS--ST mixed state. In
this case, they impose certain restrictions on the coupling constants $g_1$
and $g_2$. The sense of these restrictions is that the coupling constants in
TS and ST pairing channels must be of the same order. Clearly
that similar restrictions exist in a general case, when the
conditions of the logarithmic approximation are not fulfilled.

The results of
numerical determination of the order parameters $\De_{30}(n),\,\De_{03}(n)$
as functions of  density on the base of Eqs.  \p{15} are displayed in
Fig.~1.  It is seen that at some critical density there appear two pair $
(\pm\De_{30}(n),\De_{03}(n))$ of TS-ST solutions which thermodynamically are
indistinguishable. We conclude that two--gap TS--ST superfluid states can
arise in nuclear matter as a result of the density phase transition (at fixed
temperature) from one-gap ST superfluid state (in the model with the Skyrme
effective forces).  Note that Eqs. \p{15} have two--gap
solutions with $\De_{30}\not=0,\,\De_{03}\not=0$ in the case of the SkM,
SkM$^*$ and Ska potentials but have no such solutions for the RATP
and SkP potentials.  For the latter potentials the TS coupling constant is
considerably larger than the ST coupling constant and, hence, the conditions
for existence of two-gap solutions are broken.  Concerning the potential SkP
we remark that it has been used for the description of the pairing
correlations in highly asymmetric nuclear matter \ci{DFT} unlike to our case
where we study superfluidity of symmetric nuclear matter.

{\bf 3. Temperature behaviour of the order parameters.}
Given in the section 2 analysis relates to the case $T=0$. It is clear, that
if TS--ST states exist at $T=0$ then such states arise first at some critical
temperature $T_{tsst}$. To describe the temperature behaviour of the order
parameters and the mechanism of appearance of the new solutions it is
convenient to pass to the new independent variables $x=\De_{30}+\De_{03},\,y=
\De_{30}-\De_{03}$. With allowance for Eq. \p{12}, where we have replaced
summation by integration, the self-consistent equations in terms of
the variables $x$ and $y$ are written in the form \beq
\frac{x+y}{2g_1}=x\la(x,T)+y\la(y,T),\;
\frac{x-y}{2g_2}=x\la(x,T)-y\la(y,T),\lab{19}\eeq
$$\la(x,T)=\int_{-\th}^\th\frac{d\xi}{E}\tanh\frac{E}{2T},\;
E=\sqrt{\xi^2+x^2}$$
Excluding the variable $y$ from Eq. \p{19}, for determining the function
$x=x(T)$ we obtain the equation \beq d(x,T)\cdot d(x\cdot d(x,T),T)\equ
D(x,T)=1,\;\; d(x,T)=\frac{4g_1g_2\lambda(x,T)-g_1-g_2}{g_2-g_1}\lab{20} \eeq
It is not difficult to see that the variables $x$ and $y$ are related by
the formula $y=xd(x,T)$. Therefore, the order parameters
$\De_{30}(T),\,\De_{03}(T)$ can be found by the formulae:
\beq\De_{30}=\frac{1}{2}x(1+d(x,T)),\;\De_{03}=\frac{1}{2}x(1-d(x,T))
          \lab{21}\eeq
       One--gap solutions are obtained from Eqs. \p{21} as solutions of the
equations $d(x,T)=1$ (triplet--singlet) and $d(x,T)=-1$  (singlet--triplet)
while corresponding critical temperatures $T_{ts},\,T_{st}$ are determined
from the equations $d(0,T_{ts})=1,\,d(0,T_{st})=-1$. As follows  from
evaluations with the Skyrme forces, for all densities, where the superfluid
states exist, it holds $g_1>g_2$. Analyzing the behaviour of the function
$D(x,T)$, one can conclude that at temperatures $T>T_{ts}$ there is no
one--gap solutions; at temperatures  $T_{st}<T<T_{ts}$ there is only one
TS solution; at temperatures              $T_{tsst}<T<T_{st}$
the system is characterized by two one--gap (TS and
ST) solutions. Finally, at temperatures $T<T_{tsst}$ we have,
besides the previous solutions, two new TS--ST solutions as well.

To determine the critical temperature $T_{tsst}$, at which TS--ST solutions
arise first we use the following considerations. Clearly,  TS--ST
solutions arise as a result of branching from TS or
ST branch of solutions Eq. \p{20}. In moment, when TS--ST
solutions arise, the derivative
$D'_x(x,T)$ vanishes in the critical point  $(x_{tsst},T_{tsst})$. If
branching occurs from TS solution, i.e., $d(x,T)=1$, then
 $D'_x=d'_x(2+xd'_x)>0$ for $x>0$ (since the function $D(x,T)$ is even with
respect to  $x$, we can limit ourselves by the positive $x$; in this case
$d'_x>0$) and the equation  $D'_x=0$ has no solution. If branching occurs from
ST solution, i.e., $d(x,T)=-1$, then   $D'_x=d'_x(-2+xd'_x)$
and the critical point  $(x_{tsst},T_{tsst})$ is determined from the
equations
 \beq d(x,T)=-1,\;\; xd'_x(x,T)=2 \lab{22}\eeq
Calculation of the second derivative   $D''_{xx}(x,T)$  in the critical
point gives   $D''_{xx}(x_{tsst},T_{tsst})=0$, i.e., the mechanism of
branching of TS--ST solutions is  formation of inflection on the curve
$z=D(x,T_{tsst})$ for $x=x_{tsst}$.   All said above allows to approximate
the function $D(x,t)$ for $T<T_{st}$ in vicinity of the inflection point in
the form  \beq
D(x,T)=A(x-x_{st}(T))^3+C(T)(x-x_{st}(T))+1 \lab{23}\eeq
($x_{st}(T)$ is the singlet--triplet branch of the equation $D(x,T)=1$),
where for the coefficients
 $A\equ\frac{1}{6}D'''_{xxx}(x_{tsst},T_{tsst}),\,C(T)\equ
D'_x(x,T)|_{x=x_{st}(T)}$ it is not difficult to obtain the
expressions
$$A=\frac{1}{3}d'''_{xxx}-(d'_x)^3-\frac{3}{2}d'_xd''_{xx}-\frac{1}{2}x(d''_
{xx})^2,\; C(T)=-d'_x(x_{st}(T),T)(2-x_{st}(T)d'_x(x_{st}(T),T)).$$
 Retaining in expansion of
the function $C(T)$ the linear with respect to $T_{tsst}-T$ term, we
obtain  the formulae, describing the behaviour of TS--ST solutions
near the critical point
\beq x_{tsst}^{(\pm)}(T)=
 x_{tsst}\pm
\sqrt{\be(T_{tsst}-T)/A},\;\qu\be=2d''_{xT}-d'_T(d'_x+x_{tsst}d''_{xx});
\lab{24}\eeq
In this case, it is necessary that   $\be/A>0$.

The results of numerical determination of the order parameters
$\De_{30}(T),\,\De_{03}(T)$ on the base of Eqs. \p{20}, \p{21} are displayed
in Fig. 2. It is seen that two pairs of TS--ST solutions
$(\pm\De_{30}(T),\De_{03}(T))$ exist for the temperatures $T<T_{tsst}$ and
besides, in the critical point, in accordance with Eq. \p{21},
$\De_{30}=0,\,\De_{03}=\De_{03}^{st}$ ($\De_{03}^{st}$ being the one--gap
ST solution). We recall, that we consider the case $x>0$.
Existence of two pairs of the solutions $(\pm\De_{30}(T),\De_{03}(T))$ follows
from the fact that, if Eqs. \p{19} have as  a solution the pair of functions
$(x(T),y(T))$, then after substitution $x\ra-y,\qu y\ra-x$ we obtain the new
pair that is  a solution too. In the case $x<0$ we would be able to obtain
another two pairs of TS--ST branches $(\pm\De_{30}(T),-\De_{03}(T))$.

{\bf Conclusion.} Thus, we pointed out the possibility of arising of two-gap
unitary states in superfluid nuclear matter corresponding to the
superposition of ST and TS pairing of nucleons. The
self-consistent equations for these states differ essentially from the
equations of the BCS theory and contain the one-gap solutions (ST and TS) as
some particular cases. Analysis of the self-consistent equations at the
temperature $T=0$ in the logarithmic approximation shows that new states can
arise only under quite specific restrictions on the coupling constants, which
describe interaction of nucleons in ST and TS  pairing channels.
Since the constants of the effective
interaction depend from density, there exist such density intervals for which
either one-gap or two-gap superfluidity of nuclear matter is realized.
Calculations with the Skyrme interaction being chosen as a potential of  NN
interaction indicate that two-gap unitary states arise in nuclear matter as a
result of the phase transition in density from one-gap ST superfluid state
(results have been obtained for the Ska, SkM and SkM$^*$ potentials).
Analysis of the temperature behaviour of the order parameters shows that new
two-gap states can arise under lowering  temperature from the superfluid
state with ST pairing of nucleons as well.  Thus, we conclude that a
superfluid nucleon FL under lowering  density or temperature can undergo the
new phase transition leading to  emergence of two-gap superfluid states.
Among the other problems we note here the study of influence of asymmetry
\ci{ARS} and real bound states (deuterons) \ci{AFR,BLS} on the thermodynamic
properties of multi-gap superfluid states in nuclear matter.

The authors thank
BMBF  for partial
financial
support of this work.

\newpage
\thispagestyle{empty}
{\Large\bf  Figure captions}:

{\bf Fig. 1.} Order parameters $\De_{30},\De_{03}$ vs density of symmetric
nuclear matter at $T=0$ in the case of TS ($\De_{30}\not=0,\,\De_{03}=0$), ST
 ($\De_{30}=0,\,\De_{03}\not=0$)  and mixed TS-ST
 ($\De_{30}\not=0,\,\De_{03}\not=0$) states of paired nucleons for the
 SkM$^*$ (a), SkM (b), Ska (c) Skyrme forces; $tsst(ts_+)$, $tsst(ts_-)$
 and $tsst(st_\pm)$ are  notations for the dependences of the TS and ST order
 parameters $\De_{30}$ and $\De_{03}$ in two pair $
 (\pm\De_{30}(n),\De_{03}(n))$ of TS-ST solutions of Eq.  \p{20}. Branching
 of TS-ST solutions occurs at $n=0.049\,\mb{fm}^{-3}$ for the  SkM$^*$ potential,
 at $n=0.052\,\mb{fm}^{-3}$  for the SkM potential and
 at $n=0.064\,\mb{fm}^{-3} $ for the Ska potential.

 {\bf Fig. 2.}  Order parameters $\De_{30},\De_{03}$ vs temperature.
 Notations are the same as in Fig. 1. Calculations have been performed for
 the SkM$^*$ (a), SkM (b)  potentials at $n=0.04\,\mb{fm}^{-3}$, for the Ska
 (c) potential at $n=0.05\,\mb{fm}^{-3}$. Branching of TS-ST solutions occurs
 at $T=0.162\,\mb{MeV}$ for the SkM$^*$ potential, at $T=0.179\,
 \mb{MeV}$ for the SkM
 potential and at $T=0.203\,\mb{MeV}$ for the Ska potential.

 
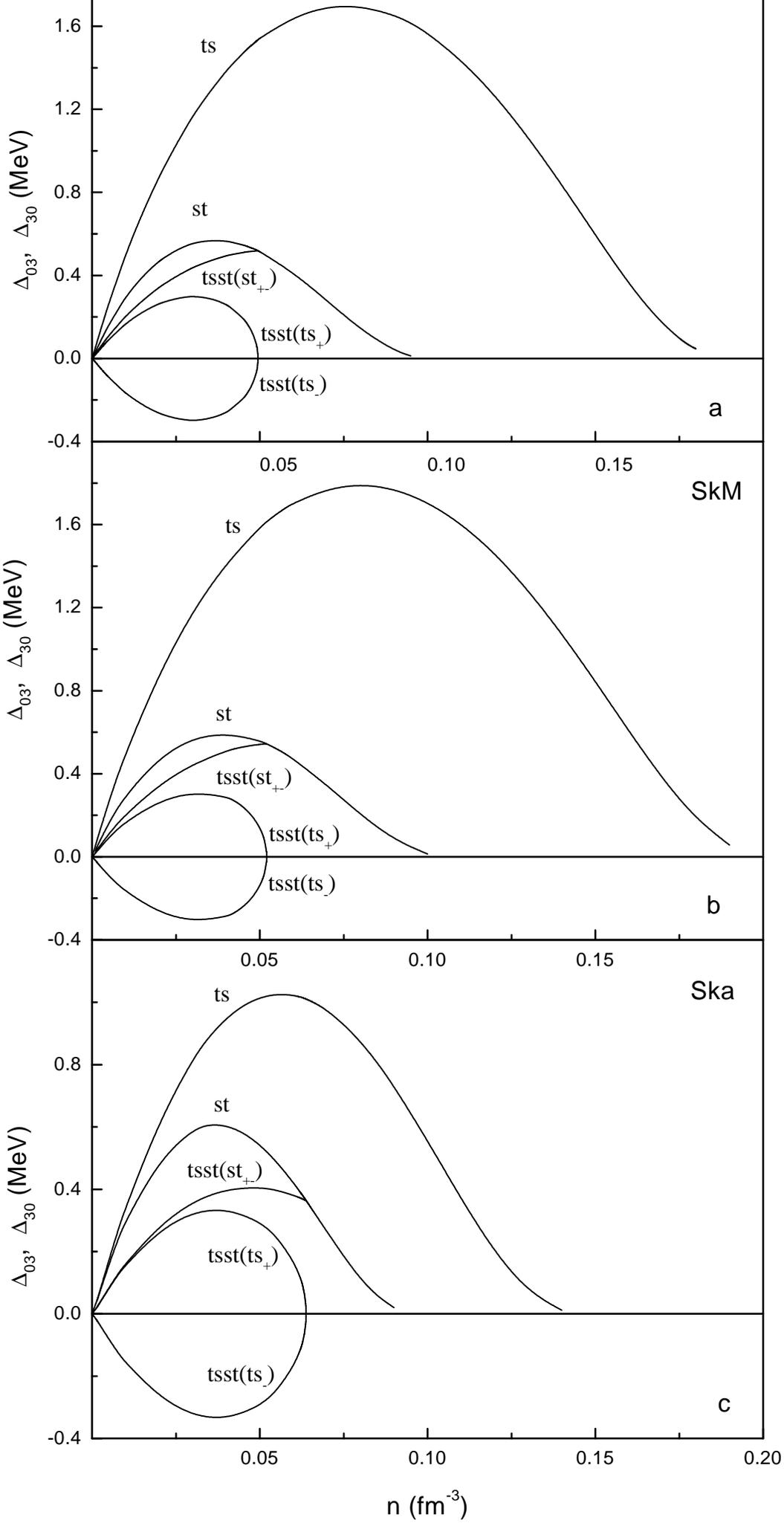
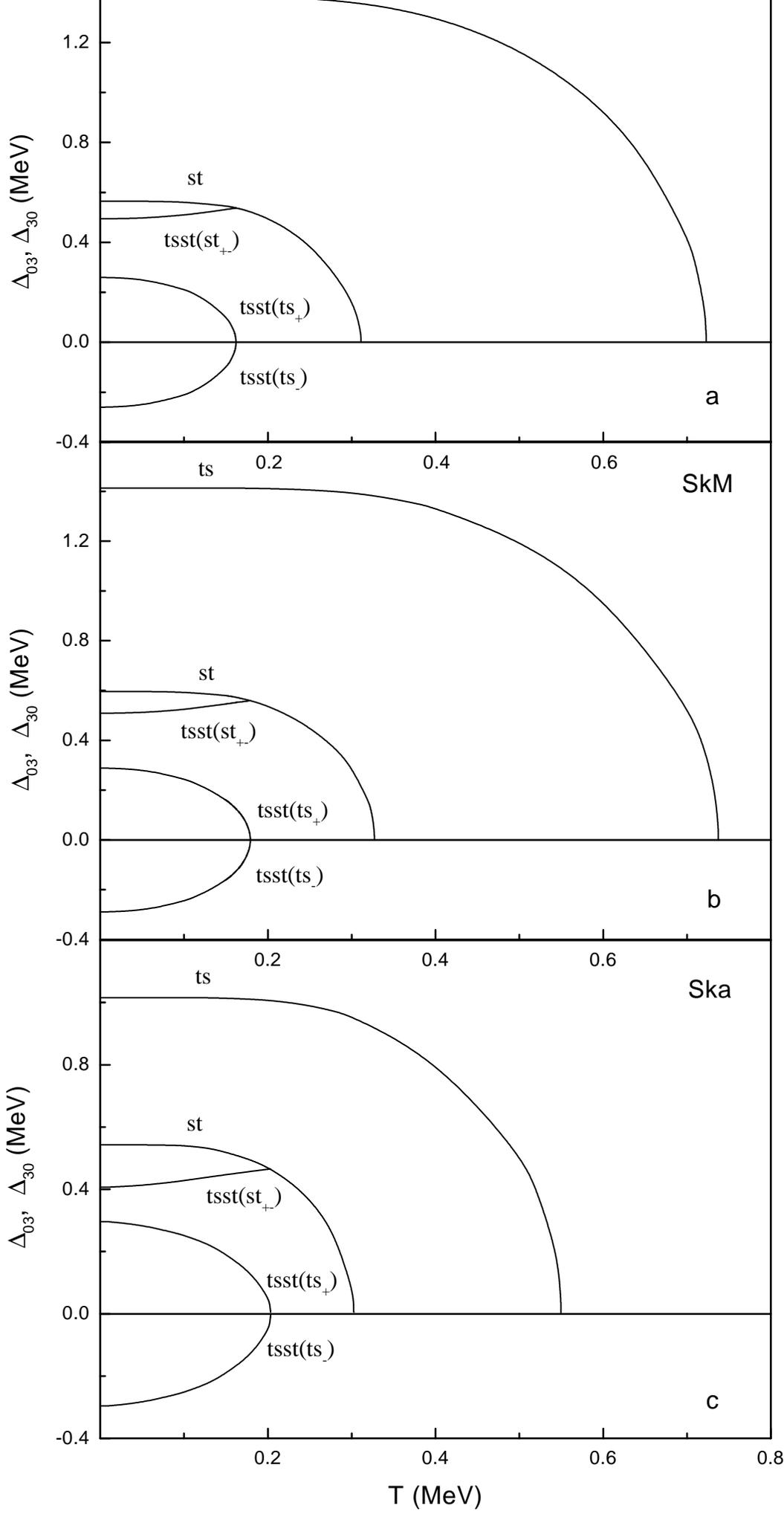
\end{document}